\documentclass[a4paper,twocolumn,accepted=2023-05-11]{quantumarticle}
\pdfoutput=1
\usepackage[utf8]{inputenc}
\usepackage[english]{babel}
\usepackage[T1]{fontenc}

\usepackage{amsmath}
\allowdisplaybreaks
\usepackage{amssymb}
\usepackage{graphicx}
\usepackage{lipsum}
\usepackage{appendix}
\usepackage{float}
\usepackage{bm}

\usepackage[italicdiff]{physics}
\usepackage[caption=false]{subfig}
\usepackage[numbers,sort&compress]{natbib}
\usepackage[colorlinks=true,allcolors=quantumviolet]{hyperref}

\begin{document}

\title{Continuous-Variable Entanglement through Central Forces:
Application to Gravity between Quantum Masses}

\author[1]{Ankit Kumar}
\orcid{0000-0003-3639-6468}
\email{kumar.ankit.vyas@gmail.com}

\author[2,3]{Tanjung Krisnanda}
\orcid{0000-0002-6360-5627}

\author[1,4]{P. Arumugam} 
\orcid{0000-0001-9624-8024}

\author[5,6]{Tomasz Paterek}
\orcid{0000-0002-8490-3156}

\affil[1]{Department of Physics,
Indian Institute of Technology Roorkee,
Roorkee 247667, India}

\affil[2]{School of Physical and Mathematical Sciences,
Nanyang Technological University, 
Singapore 637371, Singapore}

\affil[3]{Centre for Quantum Technologies,
National University of Singapore, 
Singapore 117543, Singapore}

\affil[4]{Centre for Photonics and Quantum Communication Technology,
Indian Institute of Technology Roorkee, Roorkee 247667, India}

\affil[5]{Institute of Theoretical Physics and Astrophysics,
University of Gdańsk, 80-308 Gda\'{n}sk, Poland}

\affil[6]{School of Mathematics and Physics,
Xiamen University Malaysia, 43900 Sepang, Malaysia}

\maketitle

\begin{abstract}
We describe a complete method for a precise study of gravitational interaction between two nearby quantum masses. 
Since the displacements of these masses are much smaller than the initial separation between their centers, the displacement-to-separation ratio is a natural parameter in which the gravitational potential can be expanded. 
We show that entanglement in such experiments is sensitive to initial relative momentum only when the system evolves into non-Gaussian states, i.e., when the potential is expanded at least up to the cubic term.
A pivotal role of force gradient as the dominant contributor to position-momentum correlations is demonstrated.
We establish a closed-form expression for the entanglement gain, which shows that the contribution from the cubic term is proportional to momentum and from the quartic term is proportional to momentum squared.
From a quantum information perspective, the results find applications as a momentum witness of non-Gaussian entanglement. 
Our methods are versatile and apply to any number of central interactions expanded to any order.
\end{abstract}

\section{Introduction}

Due to the weakness of gravitational coupling, all quantum experiments up to date in which gravity plays a role utilized the field of the Earth, see Refs.~\cite{PhysRevLett.4.337,doi:10.1126/science.177.4044.168,PhysRevLett.34.1472,Peters1999,Nesvizhevsky2002,PhysRevLett.118.183602} for milestone examples.
Since this field undergoes practically undetectable back-action from quantum particles, it effectively admits a classical description either in terms of a fixed background Newtonian field~\cite{PhysRevLett.34.1472,Peters1999,Nesvizhevsky2002} or as a fixed background spacetime~\cite{PhysRevLett.4.337,doi:10.1126/science.177.4044.168,PhysRevLett.118.183602}.
This argument strongly motivates theoretical and experimental research towards a demonstration of gravitation between two quantum masses, as this is one of the simplest scenarios where quantum properties of gravity could be observed.
Along this line, several proposals studied the possibility of the generation of quantum entanglement between two massive objects~\cite{PhysRevLett.119.240401,PhysRevLett.119.240402,PhysRevA.98.043811,PhysRevLett.119.120402,npjQI_6.12,JOPB_23.235501,Rijavec_2021,PhysRevLett.128.143601,PhysRevA.102.062807,PhysRevResearch.4.023087}. Our aim here is to build upon these ideas and provide a simple precision test of gravitational coupling between two nearby quantum particles. The methods we develop are generic and apply to arbitrary central interactions, including situations where many of them are present at the same time.

\begin{figure}[!b]
	\centering
\includegraphics[width=\linewidth]{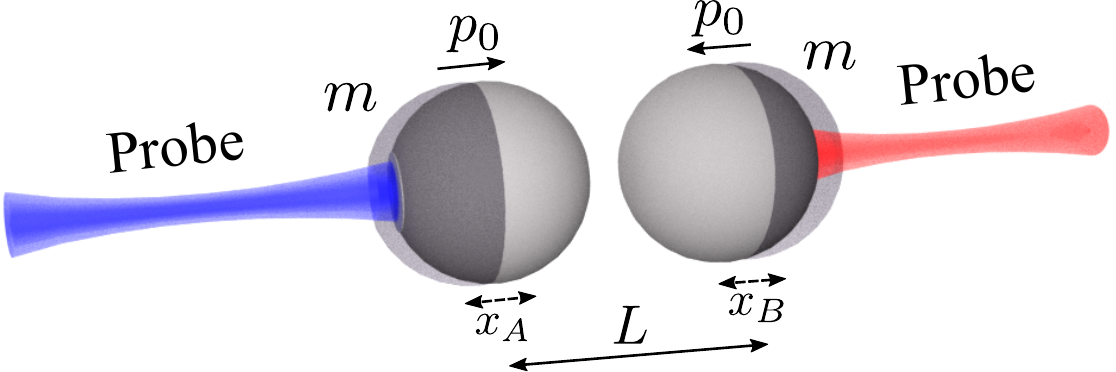}
\caption{Setup under consideration. Two spheres of mass $m$ are released from the ground state of identical harmonic traps with an equal and opposite momentum $p_0$ along the line joining their centers. The centers are initially separated by a distance $L$, and displacements from them are denoted by $x_A$ and $x_B$. After time $t$ entanglement is estimated with the help of the probing lasers.}
\label{fig:expsetup}
\end{figure}

The experiment we have in mind could be realized within the field of optomechanics~\cite{RMP_86.1391}, which already succeeded in cooling individual massive particles near their motional ground state~\cite{NJP_11.073032,Science_372.6548,Nature.478.89}, and in entangling cantilevers to light and themselves~\cite{Sciene_342.710,NatPhot.15.817,Nature.556.473}.
In such a setup the particles are separated much more than their displacements.
For example, two Osmium spheres (the densest natural material) each of mass 100 $\mu$g (radius $0.1$ mm) with an initial inter-surface distance of $0.1$ mm move by less than a nanometer within $1$ second of evolution~\cite{npjQI_6.12}, vividly illustrating the weakness of gravity. Since the situation is non-relativistic, the relevant interaction is characterised by quantum Newtonian potential.
Given that the displacements are small compared to the initial separation between the two spheres, a natural parameter in which the potential can be expanded is the displacement-to-separation ratio~\cite{npjQI_6.12,JOPB_23.235501,Datta_2021,PhysRevLett.128.110401,Roccati2022}. 
We propose to identify phenomena that can only occur if the potential is expanded to a particular order, thus witnessing the relevance of at least this order in experiments.
From this perspective, the gravitational entanglement proposals, in addition to providing clues about the quantum nature of gravity, also supply tests of the form of gravitational interaction.  
For example, entangling two initially disentangled masses requires at least the second-order term~\cite{PhysRevLett.119.240401,PhysRevLett.119.240402,npjQI_6.12}. Here we show a method that witnesses the third- and fourth-order terms and has an advantage of a simple modification of the entanglement scheme with confined particles. Hence, an experiment designed to probe gravitational entanglement can also be used to witness even weaker gravitational coupling.

Our basic idea is to push the particles towards each other as it is intuitively expected that such obtained stronger gravity will lead to higher accumulated entanglement. Yet,
we demonstrate that the quantum entanglement generated by gravitational potential truncated at the second order is \emph{insensitive} to any relative motion of the two masses. 
This is shown explicitly with an exact analytical solution for the time evolution of the corresponding covariance matrix~\cite{PRA_65.032314,PRA_70.022318,PRA_72.032334}.
Our intuition is only recovered with the potential containing at least the third-order term, i.e., when the system evolves into non-Gaussian states. 
Closed-form expressions for the amount of entanglement are established for the potential expanded to any order.
The introduced methods apply to any central interaction, even when many are present side by side. 
They also show remarkable robustness, e.g., even the impact of the fourth-order term on the non-Gaussianity quantifier and the amount of entanglement can be captured numerically despite an astonishingly weak gravitational interaction.
Moreover, the derived closed forms can be extrapolated for expansions to arbitrary order.
They are in remarkable quantitative agreement with numerical simulations, which show that the contribution from the cubic term is proportional to momentum. In contrast, the contribution from the quartic term is proportional to momentum squared. 
Accordingly, the cubic correction decreases the entanglement gain when the two particles are moving away, and the quartic correction increases the entanglement irrespective of whether they are moving towards or away from each other.

\section{Experimental setup}

\begin{figure}[!b]
\centering
\includegraphics[width=\linewidth]{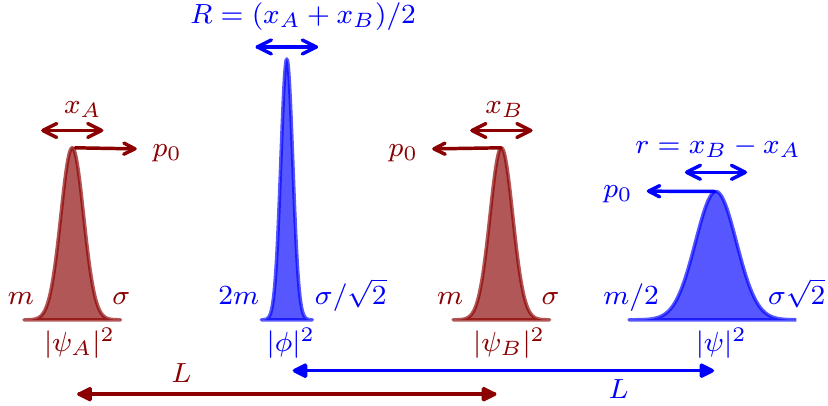}
\caption{From LAB frame to COM frame. Gaussianity of the initial state is preserved as well as the product form. The widths, however, are different in different frames as marked.}
\label{fig:TMGS_special} 
\end{figure}

Consider the setup schematically represented in Fig.~\ref{fig:expsetup}, where we introduce our notation.
The initial wave function is assumed to describe two independent masses, each in a natural Gaussian state with position spread $\sigma$: $\Psi(x_A,x_B,t=0) = \psi_A(x_A) \ \psi_B(x_B)$, where
\begin{gather}
	\psi_A(x_A) = \qty( \frac{1}{2\pi\sigma^2} )^{1/4} \exp(-\frac{x_A^2}{4\sigma^2} + i\frac{p_{0} }{\hbar} x_A ), 
\\
	\psi_B(x_B) = \qty( \frac{1}{2\pi\sigma^2} )^{1/4} \exp(-\frac{x_B^2}{4\sigma^2} - i\frac{p_{0} }{\hbar} x_B ).    
\end{gather}
Note that without loss of generality we chose the momenta to be opposite and equal. 
The Hamiltonian in the non-relativistic regime is given by
\begin{equation}
	\hat{H} = \frac{ \hat{p}_A^2}{2m}+\frac{ \hat{p}_B^2}{2m} - \frac{Gm^2}{L+( \hat{x}_B- \hat{x}_A)}.
\end{equation}

Since this is a two-body problem, it is well-known that the center-of-mass (COM) motion separates from the relative movement. 
Accordingly, we introduce the usual change of variables from the LAB frame to the COM frame:
$R = (x_A+x_B)/2$ and $r = x_B-x_A$, where $R$ and $r$ denote the respective displacements of the COM and the reduced mass from their initial average positions (see Appendix~\ref{appendix:transformations} for details). As a result, the initial wave function separates as $\Psi(x_A,x_B,t=0) = \phi(R,t=0) \ \psi(r,t=0)$,
where
\begin{gather}
	\phi(R,t=0) = \qty( \frac{1}{\pi\sigma^2} )^{1/4} \exp(-\frac{R^2}{2\sigma^2}),	\label{eq:InitialState_inCOMframe_M}
\\
	\psi(r,t=0) = \qty( \frac{1}{4\pi\sigma^2} )^{1/4} \exp(-\frac{r^2}{8\sigma^2} - i\frac{p_{0} }{\hbar} r ).
	\	\label{eq:InitialState_inCOMframe_mu}
\end{gather}
The wave functions $\phi$ and $\psi$ describe the motion of the COM and the reduced mass, respectively. Compared to the original wave packets, the COM wave packet admits a smaller width of $\sigma/\sqrt{2}$, and the reduced mass wave packet has a larger width of $\sigma\sqrt{2}$. The corresponding relations are illustrated in Fig.~\ref{fig:TMGS_special}.
In this frame the Hamiltonian decouples as
\begin{equation}
	\hat{H} =  \hat{H}_R +  \hat{H}_r = \qty(  \frac{ \hat{P}^2}{4m} ) + \qty(  \frac{ \hat{p}^2}{m} - \frac{Gm^2}{L+ \hat{r}} ),
	\label{eq:hamseparable}
\end{equation}
where $ \hat{P} = -i\hbar\partial/\partial R$ and $ \hat{p} = -i\hbar\partial/\partial r$ are the momentum operators for the COM and the reduced mass, respectively. A separable Hamiltonian implies that the two-body wave function retains its product form at all times, i.e., $\Psi(x_A,x_B,t) = \phi(R,t) \ \psi(r,t)$. 
Furthermore, the COM wave packet evolves like a free particle, i.e., its Gaussianity is preserved~\cite{JOPB_33.4447,RevModPhys.84.621,book_decoherence_Maximilian}.
The first two statistical moments characterize the quantum state fully, and for completeness, they are given in Appendix~\ref{appendix:Ehrenfest}.

The state $\psi$ evolves in the gravitational potential, which we now expand in the powers of the displacement-to-separation ratio $r/L$:
\begin{equation}
	V(\hat r) =  - \frac{Gm^2}{L+ \hat{r}} \approx  - \frac{1}{4} m \omega^2 \sum_{n=0}^{N} \frac{(-1)^n}{L^{n-2}} \hat r^n,
	\label{eq:Ham_redmass}
\end{equation}
where $N$ is the order of approximation, and we defined $\omega^2 = 4Gm/L^3$ for later convenience. 
In Appendix~\ref{appendix:Ehrenfest}, we derive exact analytical expressions for the statistical moments of $\psi$ by solving the related Ehrenfest equations in the case of $N=2$. Together with the statistical moments for the COM, these determine the covariance matrix in an exact closed-form [see Appendix~\ref{appendix:Entanglement} for the methodology].

With the inclusion of higher-order terms in the potential, i.e., $N>2$, the corresponding Ehrenfest's equations cannot be solved analytically due to the emergence of an infinite set of coupled differential equations involving ever-increasing statistical moments. 
We therefore resort to numerical methods and calculate the time evolution of $\psi$ by implementing Cayley's form of evolution operator~\cite{book_NumericalRecipies}. The numerical evaluations for $\psi$ are combined with analytical solutions for the COM to construct the covariance matrix and the two-body wave function. In order to deal with weak gravitational interaction, we improve the accuracy of Cayley's method by utilizing the five-point stencil and discretise onto a pentadiagonal Crank-Nicolson scheme, which is further solved by implementing the LU factorization techniques. The code is publicly available at Zenodo~\cite{zenodo_link}, with the corresponding documentation in Ref.~\cite{Cayley-TDSE-arXiv} where we demonstrate our superior accuracy as compared to the standard tridiagonal solutions. We also implemented a dynamic grid allocation, as described in Ref.~\cite{Quantum_5.506}, to avoid any reflections from numerical boundaries.

\section{Results}

The methodology just described returns an analytical form of the covariance matrix at time $t$ for potentials truncated at $N = 2$ and a numerical form of the two-body wave function for all $N$.
These are thereafter used for computing the entanglement between two masses [see Appendix~\ref{appendix:Entanglement} for the methodology].
In particular, we use logarithmic negativity and entropy of entanglement as entanglement quantifiers.
While logarithmic negativity is known to be a faithful entanglement quantifier for Gaussian states~\cite{PRA_65.032314,PRA_70.022318,PRA_72.032334}, we will also discuss non-Gaussian pure states and hence the inclusion of the entropy of entanglement.
We first give the results for $N=2$, emphasizing the independence of relative momentum and its origin.
Then we move to $N=3$ and demonstrate that entanglement is linearly dependent on the initial momentum.
We also analyze an indicator of non-Gaussianity (skewness) and demonstrate the precision of our methods by calculating the marginal impacts of the fourth-order term in the potential expansion. 
A methodology to obtain closed-form expressions for the entanglement gain
through potentials expanded to arbitrary order is presented.
Quantitative comparisons are made with numerical simulations for the fourth-order potential, which show that the contribution from the quartic term is proportional to relative momentum squared.

\subsection{Quadratic interactions}

\begin{figure}[!b]
\centering
\includegraphics[width=\linewidth]{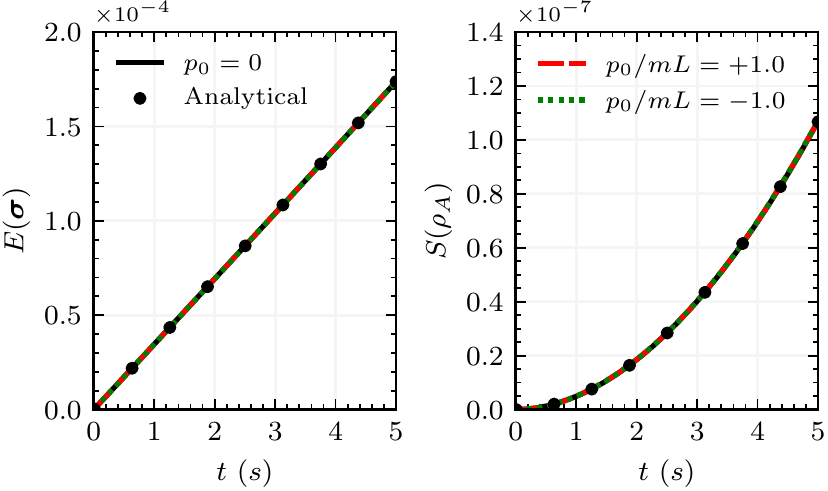}
\caption{Accumulation of entanglement with the gravitational potential truncated at the quadratic term ($N=2$). The configuration consists of identical Osmium spheres with $m = 0.25$ pg, $L=2.5$ times their radius, and $\sigma = 2.5$ nm. $p_0$ is the initial momentum. 
Analytical results are calculated with the closed form of the covariance matrix in Eq.~\!\eqref{eq:GCM_last}.
$E$ denotes the logarithmic negativity, and $S$ is the entanglement entropy. 
The values of $p_0/mL$ in the legends are written in multiples of $6.18082292 \times 10^{-3}$ s$^{-1}$.}
	\label{fig:entV2_manymom}
\end{figure}

Consider first the gravitational potential truncated at the second order. 
We obtained exact analytical forms for the independent elements of the covariance matrix $\bm{\sigma}$. The solutions simplify if they are written in terms of already introduced $\omega$ and in terms of $\omega_0 = \hbar/2m\sigma^2$, which is the frequency of harmonic trap for which the initial state is the ground state:
\begin{gather}
	\bm{\sigma}_{00}  = \frac{\hbar}{4m\omega_0} \qty[ 2+\omega_0^2t^2+\qty( 1+\frac{\omega_0^2}{\omega^2} )\sinh^2(\omega t) ],   
\nonumber \\
	\bm{\sigma}_{02}  =  \frac{\hbar}{4m\omega_0} \qty[ \omega_0^2t^2-\qty(1+\frac{\omega_0^2}{\omega^2})\sinh^2(\omega t) ],
\nonumber \\
	\bm{\sigma}_{11}   =  \frac{m\hbar\omega_0}{4} \qty[ 2+\qty( 1+\frac{\omega^2}{\omega_0^2} )\sinh^2(\omega t) ], 
\nonumber \\
	\bm{\sigma}_{13} = - \frac{m\hbar\omega_0}{4} \qty( 1+\frac{\omega^2}{\omega_0^2} ) \sinh^2(\omega t), 
\nonumber \\
	\bm{\sigma}_{01} = \frac{\hbar}{8} \qty[  2\omega_0 t + \qty( \frac{\omega_0}{\omega}+\frac{\omega}{\omega_0} )\sinh(2\omega t) ],   
\nonumber \\
	\bm{\sigma}_{03} = \frac{\hbar}{8} \qty[  2\omega_0 t - \qty( \frac{\omega_0}{\omega}+\frac{\omega}{\omega_0} )\sinh(2\omega t) ].
\label{eq:GCM_last}
\end{gather}
The logarithmic negativity for $p_0 = 0$, in the regime of $\omega \ll \omega_0$ and $\omega t \ll 1$, was already approximated to~\cite{npjQI_6.12}
\begin{equation}
	E(\bm{\sigma}) \approx -  \log_2 \sqrt{ 1+2\Omega^6t^6 -2\Omega^3t^3\sqrt{ 1 + \Omega^6t^6} },
	\label{eq:lneg_tk}
\end{equation}
where $\Omega^3 = \omega_0\omega^2 /6 \equiv \hbar G / 3\sigma^2L^3$. 
We verified that this formula indeed matches our results and emphasize that the solutions obtained in this work are \emph{exact}. Hence, they can quantify entanglement outside the demanding constraints that led to Eq.~\!\eqref{eq:lneg_tk}. An example is presented below.

The most striking feature of the covariance matrix is its insensitivity to the initial momentum $ p_0 $. Accordingly, all quantities derived from the covariance matrix, say entanglement or squeezing~\cite{Datta_2021,PhysRevA.101.063804}, are independent of the initial momentum.
In this approximation, the two initially moving masses accumulate the same amount of entanglement as when they start from rest.
Furthermore, the amount of entanglement is the same irrespective of whether the masses are moving toward or away from each other.
This is confirmed by the numerical simulations presented in Fig.~\ref{fig:entV2_manymom}.
Not only there is no momentum dependence in the dynamics of logarithmic negativity and entropy of entanglement, they also perfectly overlap with analytical results showing that our methods are reliable and consistent. We emphasize that the configurations considered here are non-relativistic. Field theory calculations imply momentum-dependent relativistic corrections to the Newtonian potential~\cite{PhysRevD.105.106028,PhysRevA.101.052110}, and accordingly, we verify that second-order quantum entanglement generated by relativistic particles is, in principle, momentum dependent. However, for the parameters in Fig.~\ref{fig:entV2_manymom}, such corrections to the Newtonian potential energy are sixteen orders of magnitude smaller, hence not discussed in this work. We also note that Eq.~\!\eqref{eq:lneg_tk} is not applicable to the configuration considered in Fig.~\ref{fig:entV2_manymom} because $\omega_0 \approx 25 \omega$.

\subsection{Relevance of force gradient}

We now move to explanations of the observed results.
Mathematically, it is clear that a non-zero force gradient across the size of the wave packet is a necessary condition for entanglement.
Without it the potential is effectively truncated at $N=1$, and the total Hamiltonian is the sum of local terms.
Physically, entanglement is caused by correlations in complementary variables, here between positions and momenta. 
Due to a force gradient, the parts of the wave packets that are closer are gravitationally attracted more than the parts which are further apart. 
Hence a moment later, different momentum is developed across different positions within the wave packets, leading to quantum entanglement.

Furthermore, assuming that the force gradient is the dominant contributor to entanglement gain explains the insensitivity to the initial momentum.
Since the potential is truncated at $N=2$, the force gradient is constant in space.
Therefore, it is irrelevant if the particle moves to a different location in the meantime; hence the initial momentum does not play a role in entanglement dynamics. Quantitatively, the force gradient is $F_2' = m \omega^2 / 2$, and therefore it fully describes entanglement in Eq.~\!\eqref{eq:lneg_tk} since now $\Omega^3 = \omega_0\omega^2/6 \equiv (\omega_0/ 3m) F_2'$.
In the following section we provide further evidence for the pivotal role of force gradient in the entanglement dynamics due to higher-order interactions.

\begin{figure}[!b]
\centering
	
\subfloat[]{\includegraphics[width=0.495\linewidth]{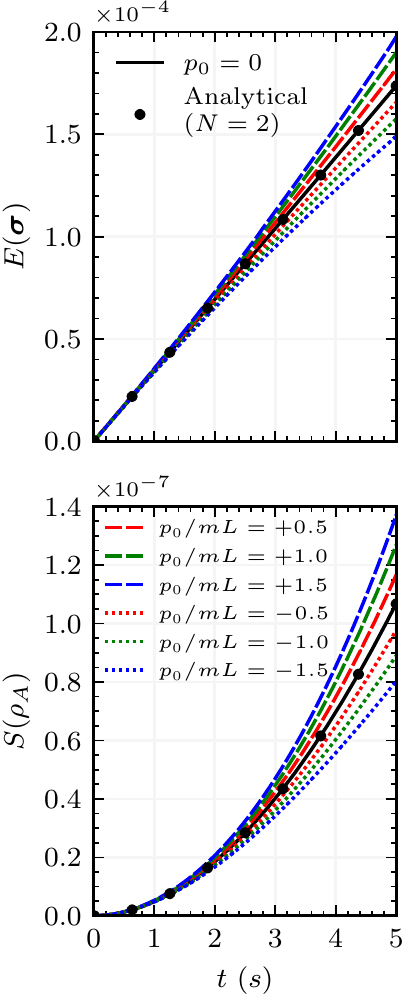}\label{fig:entV3_manymom}}
\hfill
\subfloat[]{\includegraphics[width=0.495\linewidth]{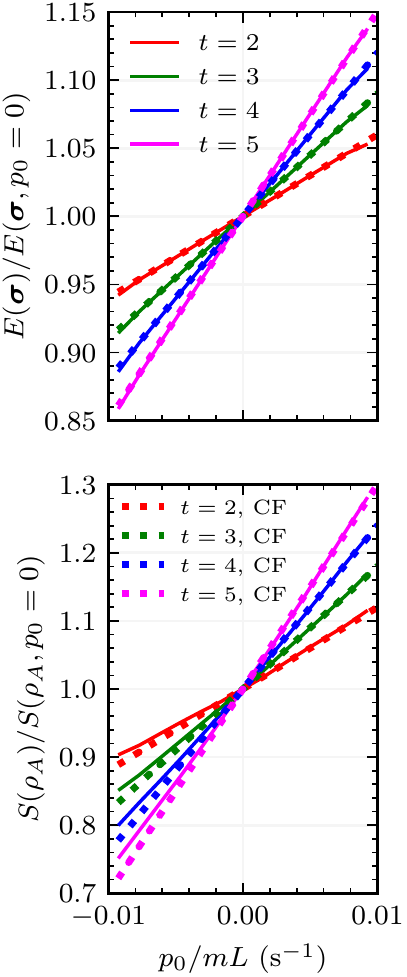}\label{fig:entV3_manytime}}
\caption{Accumulation of entanglement with gravitational potential truncated at the cubic term ($N=3$).
The same physical configuration as in Fig.~\ref{fig:entV2_manymom}.
$E$ denotes the logarithmic negativity, and $S$ is the entanglement entropy.
Panels (a) on the left show the dependence of entanglement on the initial momentum. Analytical results are calculated from the closed-form of the covariance matrix for $N=2$, and coincide with the numerical results for $N = 3$ and $p_0 = 0$.
Panels (b) on the right compare entanglement accumulated with non-zero momentum to entanglement gained from rest. The ratios show a linear dependence on the momentum and is very well approximated in the regime of positive $p_0$ (masses moving towards each other) with Eqs.~\!\eqref{eq:ent_gradestimate_S} and ~\!\eqref{eq:ent_gradestimate_E}.
The values of $p_0/mL$ in the legends are written in multiples of $6.18082292 \times 10^{-3}$ s$^{-1}$.}
\label{fig:entV3}
\end{figure}

\subsection{Higher-order interactions}

Let us continue with the working hypothesis that the force gradient is the dominant factor in entanglement dynamics.
For the cubic potential, $N=3$, the gradient is given by $F_3'( {\hat r}) = (1-3 \hat{r}/L)m\omega^2/2$ and importantly it admits a position dependence. Accordingly, entanglement should be sensitive to the initial momentum as the gradients differ at different locations.
This is indeed observed in Fig.~\ref{fig:entV3_manymom} 
for gravitational potential truncated at the cubic term. 
Furthermore, when the two masses move towards each other, $p_0 > 0$ and $\ev{\hat r} < 0$, the gradient increases, matching the growing entanglement. 
Conversely, when the masses move away, $p_0<0$ and $\ev{\hat r}>0$, the force gradient decreases, matching the slower entanglement gain. Quantitative statements can also be achieved.

Fig.~\ref{fig:entV3_manytime} shows experimentally friendly plots of the ratio of entanglement accumulated within time $t$ with non-zero initial momentum to entanglement gained from rest.
The numerically calculated linear dependence (solid lines) can be explained with closed expressions (dotted lines) that we now explain. The force gradients for the quadratic and the cubic interactions are related by the following factor: $\hat F_3'( \hat{r}) = (1-3 \hat{r}/L) F_2'$.
The average factor therefore reads
\begin{equation}
\frac{\ev{F_3'}}{F_2'} =
1-\frac{3}{L}\ev{ \hat{r}} \approx  1+ \frac{6p_0t}{mL} \equiv 1 + \epsilon_3(t),
\label{eq:ampforcegrad3}
\end{equation}
where we utilized the fact that $p_0$ is much larger than the momenta generated by gravity and the wave packet, on average, practically follows a free particle trajectory: $\ev{\hat r} \approx r_\text{cl} = -2p_0t/m$.

Fig.~\ref{fig:entV3_manymom} shows that for vanishing initial momentum, $p_0 = 0$, the entanglement obtained with cubic and quadratic potentials is practically the same.
We therefore extrapolate that entanglement for non-zero initial momentum is related to entanglement from rest
by a simple function of the conversion factor.
The plots of Fig.~\ref{fig:entV3_manytime} are fitted with
\begin{gather}
S(\rho_A) = \Big[ 1+ \epsilon_3(t)  \Big] \ S(\rho_A,p_0=0),
\label{eq:ent_gradestimate_S}
\\
E(\bm{\sigma}) = \Bigg[ 1 + \frac{1}{2} \epsilon_3(t) \Bigg] \ E(\bm{\sigma},p_0=0).
\label{eq:ent_gradestimate_E}
\end{gather}
Note that the factor of $1/2$ next to $\epsilon_3$ in the logarithmic negativity is causing a departure from the exact conversion factor between the force gradients.
These formulae are in remarkable agreement with the computational results in the regime of positive initial momentum (masses moving towards each other, the regime of experimental interest) and also work quite well for negative initial momenta.
This again affirms that the force gradient is the primary driver of gravitational entanglement.
Furthermore, these closed forms can now be used in many configurations to estimate the amplification of entanglement for a non-zero initial momentum given entanglement from rest.

\begin{figure}[!t]
\centering
\includegraphics[width=\linewidth]{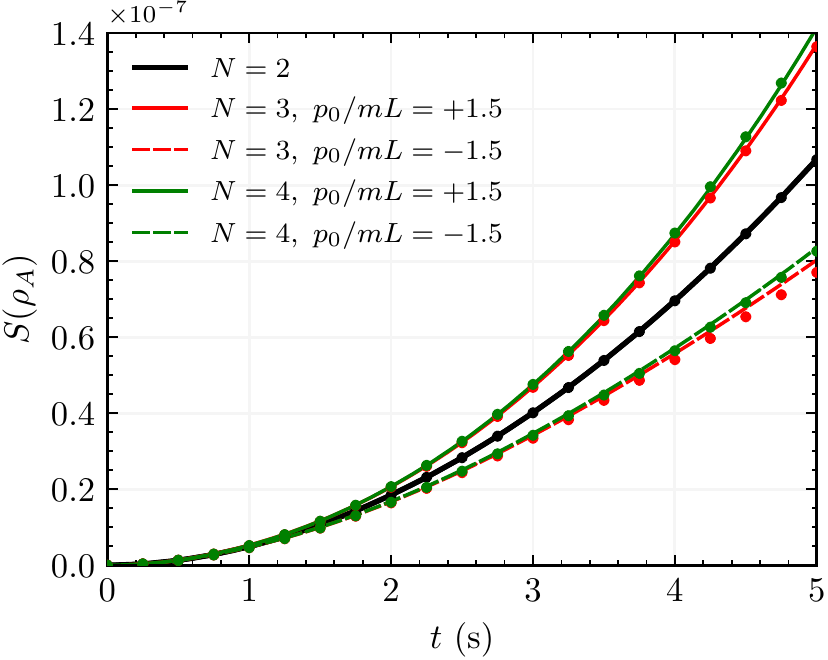}
\caption{Comparison of entanglement accumulated with the gravitational
potential expanded up to quadratic ($N = 2$), cubic ($N=3$) and quartic ($N=4$) term, respectively.
Solid lines show the results for positive momentum (masses moving toward each other), and dashed lines are for negative momentum.
The dots represent the entanglement entropy ($S$) computed with the closed formulae derived in this work.
Compared to the quadratic case, the cubic term lowers entanglement between particles that move away from each other. 
Compared to the cubic case, the quartic term adds a positive correction irrespective of the particles moving towards or away from each other.
The values of $p_0/mL$ in the legends are written in multiples of $6.18082292 \times 10^{-3}$ s$^{-1}$.}
\label{fig:entV4}
\end{figure}

For the potentials expanded to even higher-order terms, their contribution can also be incorporated with an appropriate conversion factor between the force gradients. Note that the entanglement entropy in Eq.~\!\eqref{eq:ent_gradestimate_S} is amplified in the same way as the force gradient in Eq.~\!\eqref{eq:ampforcegrad3}. Assuming that this holds for higher-order terms, the entropy amplification factor can be written as
\begin{equation}
\frac{S(\rho_A)}{S(\rho_A,p_0=0)} = 
\frac{\ev{F_N'}}{F_2'} = 1 +  \sum_{n=3}^{N} \epsilon_n(t),
\end{equation}
where the corrections for gravity-like interactions (inverse-square forces) arising due to the $n^\text{th}$ term in the potential expansion is
\begin{equation}
\epsilon_n(t) = \frac{(-1)^n}{2L^{n-2}} \ n(n-1)   \ev{\hat r^{n-2}}.
\end{equation}
Since the gravitational force between two quantum masses is weak, for the estimation of $\ev{\hat r^n}$ we 
approximate the reduced mass wave packet to be a Gaussian, with the average position following classical trajectory and the width following the free evolution:
\begin{equation}
\abs{\psi_0(r,t)}^2 \approx \frac{1}{\bm{\Delta} r_0 \sqrt{2\pi}} \exp\qty( -\frac{\qty(r-r_\text{cl})^2}{2\bm{\Delta} r_0^2} ),
\end{equation}
where $ \bm{\Delta} r_0^2 = 2 \sigma^2 \qty(1+\omega_0^2t^2)$.
With this approximation one obtains the correction terms for $n \ge 3$ as
\begin{gather}
\epsilon_n(t) = \frac{(-1)^n}{2\sqrt{\pi}L^{n-2}} \ n(n-1)    \hspace{2.5cm}
\\
\times \sum_{m=0,2,}^{n-2} {n-2 \choose m}  r_\text{cl}^{n-m-2} \ \qty( \sqrt{2}\bm{\Delta} r_0 )^m \ \Gamma \qty( \frac{m+1}{2} ),
\nonumber
\end{gather}
where $\Gamma$ is the gamma function, and the summation is only over even $m$. Note that the gravitational interaction is already included in $F_2'$, and the present estimation is for the ratio of the force gradients of different orders only, $\ev{F_N'}/ F_2'$.
Since $\epsilon_n \propto 1/L^{n-2}$, each consecutive term is diminished by a factor of $L$. Hence, a cubic order correction should be sufficient for practical applications in the near future. 
Nevertheless, one can see that the fourth-order correction to entanglement entropy is given by
\begin{equation}
\epsilon_4(t) =	24 \frac{p_0^2 t^2}{m^2 L^2}    + 12\frac{\sigma^2}{L^2} \qty(1+\omega_0^2t^2).
\end{equation}
Unlike the third-order term, which was sensitive to the direction of momentum, the fourth-order one depends on the momentum squared, leading to a positive correction in both the scenarios of masses moving towards and away from each other.
This prediction is confirmed in Fig.~\ref{fig:entV4}, where we show the entanglement accumulated with the gravitational potential expanded up to the fourth order.
The derived formulae exactly recover the entanglement gain in the regime of positive momentum, and they work quite well in the case of negative momentum.
Note that $\epsilon_4$ also depends on the position spread, hence it might be important even for stationary configurations where the wave packet undergoes a fast expansion.

\subsection{Galilean relativity and a drifting COM}

We made a change of reference frames to dissect the bipartite evolution into two independent single-particle dynamics. The first one is the free evolution of the COM, and the second one is the evolution of reduced mass in gravitational potential. Under the assumption that the two spheres are imparted with equal and opposite momentum, the COM is stationary on average. While this simplifies our theoretical framework substantially, such a configuration may be difficult to achieve in experiments. It is much easier to push one of the masses while the other is at rest. In such a case the COM moves rectilinearly with a constant velocity.

The Galilean principle of relativity demands that the laws of non-relativistic physics must be invariant in all inertial frames of reference. Consequently, the centered moments of the moving COM should evolve in the same way as for the stationary COM~\cite{AJP_36.525}. 
This is readily cross-checked as we get exactly the same correlation and variances after incorporating a non-zero momentum in the initial conditions for solving COM Ehrenfest's equations. In conclusion, a uniformly moving COM has no role in generating quantum (or, for that matter, classical) correlations. Only the relative momentum matters, and as long as it remains the same, the individual momenta can be tweaked as per convenience.

\section{Discussion}

The results presented so far could also be perceived as a simple momentum-based witness of non-Gaussianity in a quantum state. Indeed the cubic term is responsible for non-Gaussian evolution that we now quantify in more detail.

\begin{figure}[!b]
\centering	\includegraphics[width=\linewidth]{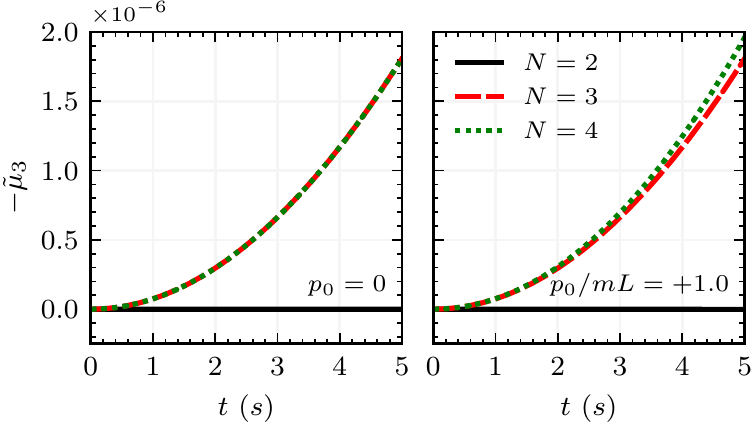}
\caption{Non-Gaussian reduced mass dynamics.
The physical situation is as in Fig.~\ref{fig:entV2_manymom}.
Skewness ($\tilde{\mu}_3$) is computed for the position space distribution. 
$N$ denotes the order of approximation, see Eq.~\!\eqref{eq:Ham_redmass}.
The left panel is for particles initially at rest, and the right panel is for masses moving toward each other. The values of $p_0/mL$ in the legends are written in multiples of $6.18082292 \times 10^{-3}$ s$^{-1}$.}
\label{fig:skewness}
\end{figure}

Fig.~\ref{fig:skewness} presents the skewness $\tilde{\mu}_3$ in the evolution of the reduced mass wave function $\psi$. While $\tilde{\mu}_3$ vanishes for $N = 2$, as it should, it rises steeply for $N = 3$.
The physical reason is clear from Fig.~\ref{fig:TMGS_special} describing the change of variables between LAB and COM frames. The left end of wave function $\psi$ is attracted towards the COM much more than the right end. Over time this makes the probability density function negatively skewed, which is indicated by $\tilde{\mu}_3 < 0$.
Just like Fig.~\ref{fig:entV4}, Fig.~\ref{fig:skewness} also demonstrates the precision of our numerical methods, which even capture marginal contributions 
of the fourth-order term.

Let us summarise the interplay between non-Gaussianity, initial momentum dependence of entanglement, and the amount of entanglement.
For the quadratic Hamiltonian the skewness of course vanishes, and we derived an exact analytical solution for the covariance matrix that shows no dependence on the initial momentum.
For the cubic Hamiltonian the skewness rises but its value is small, and within the first few seconds of the evolution the wave function is very close to a Gaussian.
Nevertheless, this is already sufficient for non-zero force gradient, and we obtained closed-form equations for the amount of entanglement that show a linear dependence on the initial momentum and do not feature skewness, see Eqs.~(\ref{eq:ampforcegrad3}) and (\ref{eq:ent_gradestimate_E}). 
Therefore, while non-zero skewness enables entanglement dependence on the initial momentum, it plays a marginal role in the amount of entanglement.
Quantitative estimation of this small contribution of skewness to the amount of entanglement is left as an open problem.

All these mutual dependencies can be seen in our data.
Fig.~\ref{fig:skewness} shows that skewness is practically the same for the two considered relative momenta. 
For the same momenta, Fig.~\ref{fig:entV3_manymom} demonstrates that entanglement entropy accumulated after 5 seconds is different by 30$\%$ and linear in initial momentum. The amount of entanglement is therefore determined by the initial momentum only.
Similarly, skewness is non-zero for initially stationary particles but entanglement dynamics with and without non-Gaussianity look practically the same.
To give quantitative values, consider the stationary configuration of two Osmium spheres with $m=1$ pg separated by a distance of $L = 2.1$ times their radius.
After an evolution for 5 seconds, the entanglement gain with cubic potential is larger than the entanglement accumulated with quadratic potential by only 
$\approx 0.001, \ 0.002$, and 
$0.003\%$, for an initial spread of $\sigma = 5.00, \ 0.50$, and $0.05$ nm, respectively.
We emphasize that the force gradient plays a pivotal role in entanglement dynamics.

We would also like to address the question of whether a simpler method for detecting the third-order coupling exists than based on measurements of entanglement. Indeed, note that solely the mean relative momentum signal could be used for such purposes.
One verifies that by truncating the potential in Eq.~\!\eqref{eq:Ham_redmass} at $N=2$, the relative momentum satisfies the condition $\ddot{\ev{p}}/\ev{p} = \omega^2$, i.e., it is time-independent. Any time dependence of this ratio reveals third-order coupling.
In cases where the center of mass is stationary, instead of the relative momentum, the local momentum of any particle could be used.

Finally, a word on decoherence effects is in place. The common decoherence mechanisms, due to thermal photons and air molecules, have already been studied in the considered setup~\cite{PhysRevLett.119.240401,PhysRevLett.119.240402,npjQI_6.12,Rijavec_2021,Datta_2021}. The experiment was found to be challenging, but the required coherence times are in principle realisable, e.g., for freely-falling particles in a high vacuum.
The calculations presented here only relax these requirements as the entanglement is improved when the two masses are pushed toward each other. For example, in the configuration considered in this work, the entanglement gain of $E \approx 1.75 \times 10^{-4}$ is relaxed from 5 seconds to 4 seconds with an initial momentum of $p_0/mL \approx +0.022$ s$^{-1}$. 
Note that an entanglement detection scheme achieving a precision of $E \sim 10^{-4}$ has recently been put forward in Ref.~\cite{Krisnanda2022_QNN}.

\section{Versatility}

While our discussions were mainly focused on gravity-induced entanglement, the methods we have presented are applicable more generally. First of all, they hold for arbitrary central interactions.
We need to expand the potential in a binomial series similar to Eq.~\!\eqref{eq:Ham_redmass} and the entanglement is characterised by the parameters $\omega$ and $\epsilon_3$. 
As we derived, for identical masses coupled via Newtonian gravity,
\begin{equation}
\omega^2 = \frac{4Gm}{L^3},		\hspace{.5cm}	\epsilon_3(t) = \frac{6p_0t}{mL}.	
\end{equation}
The general rule is quite simple. Once the potential is expanded in a binomial series of the relative displacement, the coefficient of $r^2$ is to be compared with $-m\omega^2/4$, and $\epsilon_3$ is to be calculated by comparing the force gradients as $\ev{F_3'}/F_2' = 1+\epsilon_3(t)$. One can then verify that for the Coulomb potential between charges $q_1$ and $q_2$ embedded into the masses, we obtain
\begin{equation}
\omega^2 = \frac{4q_1q_2\alpha \hbar c}{e^2mL^3},
\hspace{.5cm}
\epsilon_3(t) = \frac{6p_0t}{mL},
\end{equation}
where $\alpha$ is the fine structure constant and $e$ is the electronic charge. For the Casimir interaction between their surfaces (under proximity force approximation~\cite{PhysRevLett.99.170403}) we arrive at
\begin{equation}
\omega^2 =  \frac{\pi^3\hbar cR_0}{120m(L-2R_0)^4},
\hspace{.2cm}
\epsilon_3(t) = \frac{8p_0t}{m(L-2R_0)},
\end{equation}
where $R_0$ is the radius of each sphere.

For an arbitrary central interaction with a potential of the form
\begin{equation}
	V(x_A,x_B) =  - \frac{C}{(X+x_B-x_A)^j},
\end{equation}
we obtain
\begin{equation}
\omega^2 = \frac{2j(j+1)C}{mX^{j+2}},
\hspace{.5cm}
\epsilon_3(t) = \frac{2(j+2)p_0t}{mX} .
\end{equation}
In some situations the force is known, but solving for the potential is difficult or uncertain due to non-unique boundary conditions. In those cases the general rule would be to expand the force in a binomial series and compare the coefficient of $r$ with $m\omega^2/2$. For example, if the force is of the form
\begin{equation}
	F(x_A,x_B) =  - \frac{C}{(X+x_B-x_A)^j}  ,
\end{equation}
one arrives at
\begin{equation}
\omega^2 = \frac{2jC \ }{mX^{j+1}},
\hspace{.5cm}
\epsilon_3(t) = \frac{2(j+1)p_0t}{mX} .
\end{equation}
Note that the functional forms of $\epsilon_3$ in this section are valid only for weak interactions. 
For stronger potentials one has to take a step back and use $\epsilon_3 = -3\ev{\hat r}/L$, where  $\ev{\hat r}$ has to be approximated either analytically or numerically.

Furthermore, the methods established also work for multiple central forces acting simultaneously. If we write the interaction as a sum
\begin{equation}
V = \sum_k  V_k(x_B-x_A) , 	\hspace{.25cm}	F = \sum_k  F_k(x_B-x_A) ,
\end{equation}
the total $\omega$ characterising the Gaussian covariance matrix is given by a Pythagoras-like theorem, and the equivalent $\epsilon_3$ governing the entanglement amplification due to the cubic-order term is calculated to be a weighted sum:
\begin{equation}
\omega^2 = \sum_k \omega^2_k,
\hspace{.5cm}
\epsilon_3(t) = \frac{1}{\omega^2} \sum_k \omega_k^2 \epsilon_{k3}(t) .
\end{equation}
where $\omega_k$ and $\epsilon_{k3}(t)$  characterise the individual interactions. This is particularly useful from an experimental point of view as, in practice, it might be difficult to screen all interactions except gravity. For example, the gravitational and the Casimir interaction will likely act side by side.

\section{Conclusions}

We have shown that experiments aimed at observing gravitational entanglement can also be used as precision tests of gravitational coupling.
In particular, entanglement dependence on the relative momentum of interacting particles indicates a coupling which is higher than quadratic. 
Furthermore, for the potential expanded to the cubic term, the amount of entanglement accumulated in a fixed time interval grows linearly with the relative momentum when the particles are pushed toward each other.
We presented a closed expression for the amount of entanglement as a function of relative momentum based on the derived exact covariance matrix for Gaussian dynamics extended to higher-order couplings. 
The methods introduced apply to arbitrary central interactions, even when many are present side by side.

\begin{acknowledgements}
This work is jointly supported by 
(i) NAWA, Poland, via project  PPN/PPO/2018/1/00007/U/00001, 
(ii) XMUM, Malaysia, via project XMUMRF/2022-C10/IPHY/0002, and
(iii) DORA office of IIT Roorkee, India.
A.K. thanks the IIT Roorkee Heritage Foundation, USA, for the `Pledge a Dream' grant.
Discussions with Andy Chia (CQT-NUS, Singapore) are acknowledged.
T.K. thanks Timothy Liew for their hospitality at NTU, Singapore, and Yvonne Gao for their hospitality at NUS, Singapore. 
We acknowledge the National Supercomputing Mission (NSM) for providing computing resources of `PARAM Ganga’ at IIT Roorkee, India, which is implemented by C-DAC and supported by MeitY and DST, Govt.~of India. 
The authors gratefully thank the reviewers for their comments and suggestions which have significantly improved the presentation of this article.
\end{acknowledgements}

\appendix
\appendixpage

\section{Coordinate transformations}
\label{appendix:transformations}

The time-dependent Schr\"{o}dinger equation (TDSE) for two identical particles $A$ and $B$ is given by
\begin{eqnarray}
	\qty( -\frac{\hbar^2}{2m} \pdv[2]{x_A}  -\frac{\hbar^2}{2m} \pdv[2]{x_B} + V(\hat x_A,\hat x_B)  ) \Psi(x_A,x_B,t) 
\nonumber	\\
= i\hbar\pdv{t} \Psi(x_A,x_B,t), \hspace*{.5cm}
\end{eqnarray}
where $\hat p_A = -i\hbar\partial/\partial x_A$ and $\hat p_B = -i\hbar\partial/\partial x_B$ are their respective momentum operators. 
Coordinate transformations between the LAB the COM frame of reference are defined by
\begin{equation}
R = \frac{x_A+x_B}{2},
\hspace{.5cm}
r = x_B-x_A,
\end{equation}
where $R$ and $r$ are the displacements of the COM (mass $2m$) and the reduced mass (mass $m/2$), respectively.
One can take time derivatives to arrive at their respective momenta as
\begin{equation}
P = p_A+p_B,
\hspace{.5cm}
p = \frac{p_B-p_A}{2} .
\end{equation}
Accordingly, the inverse transformations are
\begin{equation}
x_A(x_B) = R -\!(\!+\!) \ \frac{r}{2},
\hspace{.5cm}
p_A(p_B) = \frac{P}{2} -\!(\!+\!) \ p.
\end{equation}
The displacements $x_A$ and $x_B$ are functions of $R$ and $r$, and the rules of differentiation imply
\begin{equation}
    \pdv[2]{x_A}\qty(\pdv[2]{x_B}) = \frac{1}{4} \pdv[2]{R} + \pdv[2]{r} -\!(\!+\!) \pdv{}{R}{r},
\end{equation}
which means that the kinetic energy is equivalent to
\begin{equation}
   - \frac{\hbar^2}{2m}\pdv[2]{x_A} - \frac{\hbar^2}{2m}\pdv[2]{x_B} = 
 - \frac{\hbar^2}{4m} \pdv[2]{R}   - \frac{\hbar^2}{m} \pdv[2]{r}.
\end{equation}
For central potentials $V(x_A,x_B) = V(x_B-x_A) \equiv V(r)$, and hence the TDSE is transformed to
\begin{eqnarray}
	\qty( - \frac{\hbar^2}{4m} \pdv[2]{R}   - \frac{\hbar^2}{m} \pdv[2]{r} + V(\hat r)  ) \Psi(x_A,x_B,t)
\nonumber	\\
=  i\hbar\pdv{t}\Psi(x_A,x_B,t).
	\label{eq:TDSE_in_COM}
\end{eqnarray}
In this work the initial wave function is $\Psi(x_A,x_B,t=0) = \phi(R,t=0) \ \psi(r,t=0)$, and the separation of variables in Eq.~\!\eqref{eq:TDSE_in_COM} ensures that the product form is maintained at all times. Accordingly, the problem decouples into two independent single-particle TDSEs:
\begin{gather}
-\frac{\hbar^2}{4m} \pdv[2]{R} \phi(R,t)  =  i\hbar\pdv{t}\phi(R,t),
\\
\qty( -\frac{\hbar^2}{m} \pdv[2]{r} + V(\hat r) ) \psi(r,t) = i\hbar\pdv{t}\psi(r,t),
\end{gather}
where $\hat P = -i\hbar\partial/\partial R$ and $\hat p = -i\hbar\partial/\partial r$ can now be identified as the momentum operators for the COM and the reduced mass, respectively. The COM is a free particle (but not a plane wave), and the reduced mass undergoes an evolution under the influence of the interaction. The bipartite wave function is simply the product of the two wave functions
\begin{equation}
\Psi(x_A,x_B,t) = \phi \qty( \frac{x_A+x_B}{2},t  ) \ \psi \qty( x_B-x_A,t ).
\label{eq:TBWF_LAB2COM_relation}
\end{equation}

\section{The Ehrenfest dynamics}
\label{appendix:Ehrenfest}

Ehrenfest's theorem relates the time derivative of the expectation value of an operator $ \hat A$ to the expectation of its commutator with the Hamiltonian $\hat H$:
\begin{equation}
\dv{t}\ev{\hat A} = \frac{1}{i\hbar} \ev{ \comm{\hat A}{\hat H} } + \ev{\pdv{\hat A}{t}}.
\end{equation}
In this work we focus on the noiseless dynamics, i.e., when the system is isolated from the environment.
Accordingly, the COM and the reduced mass evolve exclusively into pure states, with their covariance matrices satisfying $\text{Det}(\bm\sigma_R) 
= \text{Det}(\bm\sigma_r) 
= (\hbar/2)^2$~\cite{Serafini_2003}.

\subsection{Evolution of the COM}

The COM evolution corresponds to the free expansion of a Gaussian wave packet: $\hat H_R =  \hat P^2/4m$.
The initial is characterised by $\ev{\hat R} = 0$, $\ev{\hat P} = 0$, $\ev{ \acomm{\hat R}{\hat P} } = 0$, $\ev{\hat R^2} = \sigma^2/2$, and $\ev{\hat P^2} = \hbar^2/2\sigma^2$ [see Fig.~\ref{fig:TMGS_special} or Eq.~\!\eqref{eq:InitialState_inCOMframe_M}].
The solution to the corresponding Ehrenfest's differential equations for the first two statistical moments,
\begin{gather}
\dv{t} \ev{\hat R} 
=	\frac{1}{2m}\ev{\hat P},
	\nonumber	\\
\dv{t} \ev{\hat P} = 0,	
\nonumber	\\
\dv{t} \ev{\acomm{\hat R}{\hat P}}   = \frac{1}{m} \ev{P^2},
	\nonumber	\\
\dv{t} \ev{\hat R^2} = \frac{1}{2m} \ev{\acomm{\hat R}{\hat P}},
	\nonumber	\\
\dv{t} \ev{\hat P^2} = 0,
\end{gather}
imply
\begin{gather}
\bm{\Delta} R^2 =  \frac{1}{2}\sigma^2 (1+\omega_0^2t^2),	
\nonumber \\
\bm{\Delta} P^2 =  \frac{\hbar^2}{2\sigma^2},	
\nonumber \\
\textbf{Cov}( {R}, {P})  = \frac{1}{2}\hbar\omega_0 t.
\end{gather}
Alternatively, one arrives at the same results by utilizing the functional form of the wave function~\cite{AJP_36.525}:
\begin{equation}
\phi(R,t) = \frac{1}{\sqrt{\sigma(1+i\omega_0t)\sqrt{\pi}}} \exp\qty( -\frac{R^2}{2\sigma^2(1+i\omega_0t)} ).
\label{eq:TDWF_COM}
\end{equation}

\subsection{Evolution of the reduced mass}

The reduced mass Hamiltonian can be represented by a binomial series as in Eq.~\!\eqref{eq:Ham_redmass}. The initial state is characterised by $\ev{\hat r} = 0$, $\ev{\hat p} = p_0$, $\ev{ \acomm{\hat r}{\hat p} }  = 0$, $\ev{\hat r^2} = 2\sigma^2$, and $\ev{\hat p^2} = p_0^2+\hbar^2/8\sigma^2$ [see Fig.~\ref{fig:TMGS_special} or Eq.~\!\eqref{eq:InitialState_inCOMframe_mu}]. For $N=2$, i.e., a quadratic Hamiltonian,
\begin{equation}
\hat{H}_r = \frac{ \hat{p}^2}{m} - \frac{1}{4} m \omega^2 \qty( L^2 - Lr + r^2 ),
\end{equation}
the relevant Ehrenfest's coupled differential equations for the first two statistical moments are given by
\begin{gather}
    	\dv{t} \ev{\hat r}
     = 	\frac{2}{m}\ev{\hat p},
	\nonumber	\\
	\dv{t} \ev{\hat p} 
 =  -\frac{1}{4}m\omega^2L + \frac{1}{2}m\omega^2\ev{\hat r},	
	\nonumber	\\
	\dv{t} \ev{ \acomm{\hat r}{\hat p} } 
 =	\frac{4}{m}\ev{\hat  p^2} -\frac{1}{2}m\omega^2L\ev{\hat r} + m\omega^2\ev{\hat r^2},
    \nonumber	\\
	\dv{t} \ev{r^2} 
 =	 \frac{2}{m}
	\ev{ \acomm{\hat r}{\hat p} } ,
		\nonumber	\\
	\dv{t} \ev{\hat p^2} 
 =  \frac{1}{2}m\omega^2 \ev{ \acomm{\hat r}{\hat p} } - \frac{1}{2}m\omega^2L\ev{\hat p},
\end{gather}
Their exact analytical solution reads
\begin{eqnarray}
\ev{ \hat{r}}  &=&  \frac{1}{2}L \Big( 1 -\cosh(\omega t) \Big) -  \frac{2p_0}{m\omega} \sinh(\omega t),
\nonumber	\\	\nonumber	\\
	\ev{ \hat{p}}  &=& - p_0\cosh(\omega t) - \frac{1}{4} m\omega L \sinh(\omega t),
\nonumber	\\	\nonumber	\\
\ev{ \acomm{\hat r}{\hat p} }  &=& Lp_0 \Big( \cosh(2\omega t) - \cosh(\omega t) \Big)	
\nonumber	\\
&& + \frac{2}{m\omega} \qty( p_0^2+\frac{\hbar^2}{8\sigma^2}+ \frac{1}{2}m^2\omega^2\sigma^2 ) \sinh(2\omega t)
\nonumber	\\
&& + \frac{1}{8}m\omega L^2 \Big( \sinh(2\omega t) - 2\sinh(\omega t) \Big), 
\nonumber	\\	\nonumber	\\
\ev{ \hat{r}^2 }  &=& 2\sigma^2 \Big( 1+\sinh[2](\omega t) \Big)	
\nonumber	\\
&& + \frac{1}{8}L^2 \Big( 3 + \cosh(2\omega t)-4\cosh(\omega t) \Big)
\nonumber	\\
&& + \frac{Lp_0}{m\omega} \Big( \sinh(2\omega t) - 2\sinh(\omega t) \Big) 	
\nonumber	\\
&& + \frac{4}{m^2\omega^2} \qty( p_0^2+\frac{\hbar^2}{8\sigma^2} ) \sinh[2](\omega t),
\nonumber	\\	\nonumber	\\
\ev{ \hat{p}^2}  &=& \qty(p_0^2+\frac{\hbar^2}{8\sigma^2} ) \Big( 1+\sinh[2](\omega t) \Big)
\nonumber	\\
&& + \frac{1}{4}m^2\omega^2 \qty( 2\sigma^2+\frac{1}{4}L^2 ) \sinh[2](\omega t)
\nonumber	\\
&& + \frac{1}{4}m\omega Lp_0\sinh(2\omega t) ,
\label{eq:relmass_p2}
\end{eqnarray}
which implies that variances and the correlation are
\begin{gather}
\bm{\Delta} r^2 = 2\sigma^2 \qty( \cosh[2](\omega t) + \frac{\omega_0^2}{\omega^2} \sinh[2](\omega t) ),   
\nonumber \\
\bm{\Delta} p^2  = \frac{\hbar^2}{8\sigma^2} \qty( \cosh[2](\omega t) +  \frac{\omega^2}{\omega_0^2} \sinh[2](\omega t) ), 
\nonumber \\
\textbf{Cov}( {r}, {p}) = \frac{\hbar}{4} \qty( \frac{\omega_0}{\omega}+\frac{\omega}{\omega_0}  ) \sinh(2\omega t).
\end{gather}

\section{Quantification of entanglement}
\label{appendix:Entanglement} 

We have employed the formalism based on the covariance matrix to quantify entanglement gain via logarithmic negativity and additionally used the density matrix to compute the entropy of entanglement.

\subsection{Covariance matrix}

The covariance matrix formalism is based on the first two statistical moments of a quantum state. Given a bipartite system $AB$ with $ \hat u = ( \hat x_A, \hat p_A, \hat x_B, \hat p_B)^T$, the covariance matrix is defined as~\cite{PRA_65.032314,PRA_70.022318,PRA_72.032334}:
\begin{equation}
	\bm{\sigma}_{jk}  = \frac{1}{2} \ev{ \acomm{\hat u_j}{\hat u_k} }  -  \ev{ \hat u_j }\ev{ \hat u_k } .
\end{equation}
In the block form we can write
\begin{equation}
\bm{\sigma} \equiv \mqty(
	\bm{\alpha} & \bm{\gamma} \\
	\bm{\gamma}^T & \bm{\beta}
	),
\end{equation}
where $\bm{\alpha}(\bm{\beta}$) contains the local mode correlation for $A(B)$, and $\bm{\gamma}$ describes the intermodal correlation. 
In our setting the local modes are identical, i.e., $\bm{\alpha} = \bm{\beta}$, and a coordinate change to the COM frame implies
\begin{gather}
\bm{\sigma}_{00}(\bm{\sigma}_{02})   = \bm{\Delta} R^2 +\!(\!-\!) \ \frac{1}{4} \bm{\Delta} r^2,   
\nonumber \\
\bm{\sigma}_{11}(\bm{\sigma}_{13})  =  \frac{1}{4} \bm{\Delta} P^2 +\!(\!-\!) \ \bm{\Delta} p^2,
\nonumber \\
\bm{\sigma}_{01}(\bm{\sigma}_{03}) =  \frac{1}{2} \textbf{Cov}( {R}, {P})  +\!(\!-\!) \ \frac{1}{2} \textbf{Cov}( {r}, {p}).
\end{gather}
Given the symmetry of the problem we have $\bm{\sigma}_{22} = \bm{\sigma}_{00}$, $\bm{\sigma}_{33} = \bm{\sigma}_{11}$, $\bm{\sigma}_{23} = \bm{\sigma}_{01}$, $\bm{\sigma}_{12} = \bm{\sigma}_{03}$.
The rest of the elements are constrained due to the symmetry property $\bm{\sigma}_{jk} = \bm{\sigma}_{kj}$.

\subsection{Logarithmic negativity}

The negativity of partially transposed density matrix is a necessary and sufficient condition for entanglement in two--mode Gaussian states~\cite{PRL_84.2726}. As a result of partial transposition, the covariance matrix is transformed to $\tilde{\bm{\sigma}}$, which differs from $\bm{\sigma}$ by a sign-flip of $\text{Det} (\bm{\gamma})$~\cite{PRA_72.032334}.
The symplectic eigenvalues of the covariance matrix, $\tilde{\nu}_{\pm}(\bm{\sigma})$, are given by~\cite{PRA_70.022318,PRA_65.032314}
\begin{equation}
2\tilde{\nu}^2_{\pm}(\bm{\sigma}) =  
 \tilde{\Sigma}(\bm{\sigma}) \pm \sqrt{\tilde{\Sigma}^2(\bm{\sigma}) - 4 \ \text{Det}\qty( \bm{\sigma} ) }  ,
\end{equation}
where $\tilde{\Sigma}(\bm{\sigma}) = \text{Det} (\bm{\alpha}) + \text{Det} ( \bm{\beta}) - 2 \ \text{Det} ( \bm{\gamma} )$. For the symmetric problem as considered in this work the local modes  $ \bm{\alpha}$ and $ \bm{\beta}$ are identical, and hence $\tilde{\Sigma}(\bm{\sigma}) = 2 \ \qty[ \text{Det} (\bm{\alpha}) - \text{Det} ( \bm{\gamma} ) ] $. Entanglement is quantified by the minimum symplectic eigenvalue via logarithmic negativity:
\begin{equation}
E(\bm{\sigma}) = \max \Bigg[ 0, \ -\log_2\qty(  \frac{\tilde\nu_-(\bm{\sigma})}{\hbar/2} ) \Bigg].
\label{eq:E_from_covmat}
\end{equation}

\subsection{Entropy of entanglement}

For a pure bipartite system described by a density matrix $\rho_{AB}$, the entanglement entropy is defined as the von Neumann entropy for any one of the subsystems, e.g., $S(\rho_A) = - \Tr\qty[ \rho_A\log_2(\rho_A) ]$, where $\rho_A = \Tr_B\qty( \rho_{AB} )$ is the reduced density matrix for subsystem $A$. In order to calculate $S(\rho_A)$ we start with the two-body wave function of Eq.~\!\eqref{eq:TBWF_LAB2COM_relation}. The COM wave function $\phi(R,t)$ is derived analytically in Eq.~\!\eqref{eq:TDWF_COM}, and we calculate $\psi(r,t)$ numerically by implementing the improved Cayley's propagator~\cite{zenodo_link}. Once this is available at a given time $t$, we perform a singular value decomposition~\cite{doi:10.1119/1.17904,thesis_pyqentangle}:
\begin{equation}
\Psi(x_A,x_B,t) = \sum_{j} \sqrt{\lambda_j(t)} \ \chi^{(A)}_j(x_A,t) \ \chi^{(B)}_j(x_B,t) ,
\end{equation}
where $\qty{\chi^{(A)}_j}$ and $\qty{\chi^{(B)}_j}$ are orthonormal states in subsystems $A$ and $B$, respectively, and $\lambda_j$ are the Schmidt coefficients. A numerical implementation utilizes the algorithms in Google TensorNetwork~\cite{arXiv:1905.01330,TensorNetwork-GitHub} with the help of open source libraries hosted at GitHub ~\cite{pyqentangle-GitHub}. 
Note that the total number of Schmidt coefficients is not fixed.
At any given time, the number is dynamically increased until the norm bipartite wave function is recovered up to seven decimal places, i.e., until $1-\abs{\Psi(x_A,x_B,t)}^2 = 1-\sum_j \lambda_j(t) \lesssim 10^{-7}$.
With this decomposition, the entanglement entropy reduces to
\begin{equation}
S(\rho_A)  = -\sum_{j} \lambda_j \log_2(\lambda_j).
\end{equation}
In the case of Gaussian evolution, the entanglement entropy is calculable using the covariance matrix~\cite{Serafini_2003}:
\begin{equation}
S(\bm{\alpha}) = f\qty( \frac{1}{\hbar}\sqrt{\text{Det} (\bm{\alpha})} ) \equiv S(\rho_A) , 
\label{eq:S_from_covmat}
\end{equation}
where
\begin{equation}
    f(x) = \qty(x\!+\frac{1}{2})\log_2\qty(x\!+\frac{1}{2})
 - \qty(x\!-\frac{1}{2})\log_2\qty(x\!-\frac{1}{2}).
\end{equation}

\section{Numerical details}

Numerical calculations are performed in natural units of $c=1$, hence the conversion constant $\hbar c = 197.3269804$ keV pm. The density of Osmium is $22.5872$ g/cm$^3$. An error analysis implies that, in the numerical time evolution of the reduced mass wave function, a grid size of $\lesssim 0.2$ pm with a time step of $\lesssim 10 \ \mu$s is required to maintain accuracy in the extreme cases of the largest momentum considered in this work. 
Accordingly, we set a grid size of $0.1$ pm and a time step of $5 \ \mu$s throughout this work.
Note that the first term in the gravitational potential of Eq.~\!\eqref{eq:Ham_redmass} is just a constant energy offset, which only contributes to an irrelevant global phase in the quantum dynamics. Moreover, this term is the most prominent of all magnitude-wise. We have therefore ignored it in numerical simulations to maximally utilize the computer precision for the (relevant) higher-order terms.

\bibliographystyle{quantum}
\bibliography{main}

\end{document}